\documentclass{iau}
\usepackage{graphicx}
\newcommand{\kms}{km\,s$^{-1}$}
\newcommand{\arcsec}{$^{\prime\prime}$}

\title[AGN Feeding and Feedback] 
{Feeding and Feedback in nearby AGN -- \\ Comparison with the Milky Way center}

\author[Storchi-Bergmann]   
{Thaisa Storchi-Bergmann$^1$}

\affiliation{$^1$Instituto de F\'{i}sica, Universidade Federal do Rio
Grande do Sul \\ Campus do Vale, CP 15051, 91501-970 Porto Alegre RS, Brazil
\\ email: {\tt thaisa@ufrgs.br}}

\pubyear{2014}
\volume{303}  
\pagerange{119--126}
\jname{Feeding and feedback in a normal galactic nucleus}
\editors{Lor\'ant Sjouwerman, J\"urgen Ott \& Cornelia Lang, eds.}

\begin{document}

\maketitle

\begin{abstract}
I discuss feeding and feedback processes observed in the inner few hundred parsecs of nearby active galaxies using integral field spectroscopy
at spatial resolutions of a few to tens of parsecs. Signatures of {\it feedback} include outflows from the nucleus with velocities ranging from 200 to 1000\,\kms, with mass outflow rates between 0.5 and a few M$_\odot$\,yr$^{-1}$. Signatures of  {\it feeding} include the observation of gas inflows along nuclear spirals and filaments, with velocities ranging from 50 to 100\,\kms\ and mass flow rates from 0.1 to $\sim$\,1\,M$_\odot$\,yr$^{-1}$. These rates  are 2--3 orders of magnitude larger than the mass accretion rate to the supermassive black hole (SMBH). These inflows can thus lead, during less than one activity cycle, to the accumulation of enough gas in the inner few hundred parsecs, to trigger the formation of new stars, leading to the growth of the galaxy bulge. Young to intermediate age stars have indeed been found in circumnuclear rings around a number of Active Galactic Nuclei (AGN). In particular, one of these rings, with radius of $\approx$\,100\,pc is observed in the Seyfert 2 galaxy NGC\,1068, and is associated to an off-centered molecular ring, very similar to that observed in the Milky Way (MW). 
On the basis of an evolutionary scenario in which gas falling into the nuclear region triggers star formation followed by the triggering of nuclear activity, we speculate that, in the case  of the MW, molecular gas has already accumulated within the inner $\approx$\,100\,pc to trigger the formation of new stars, as supported by the presence of blue stars close to the galactic center. A possible increase in the star-formation rate in the nuclear region will then be followed, probably tens of millions of years later, by the triggering of nuclear activity in Sgr A*.

\end{abstract}

\firstsection 
\section{Introduction}

In the present paradigm of galaxy formation and evolution, supermassive black holes (hereafter SMBHs) are present in most galaxy bulges (\cite[Ferrarese \& Ford 2005]{ff05}). These SMBH evolve, together with the host galaxy (\cite[Di Matteo et al. 2008]{dimatteo08}; \cite[Kormendy \& Ho 2013]{kh13}) via episodes of mass accretion to the nuclear region, which may lead both to the growth of the galaxy bulge via the formation of new stars as well as to the phenomenon of nuclear activity via mass accretion to the SMBH (feeding).  The nuclear activity, once triggered, drives mechanical and radiative feedback (\cite[Hopkins et al. 2005]{hopkins05}; \cite[Di Matteo et al. 2005]{dimatteo05}), which influences the evolution of their host galaxies. Feeding and feedback processes which occur in Active Galactic Nuclei (AGN) thus constrain the co-evolution of galaxies and SMBHs, but their implementation have been simplistic (e.g. \cite[Springel et al. 2005]{springel05}; \cite[Croton et al. 2006]{croton06}; \cite[Sommerville et al. 2008]{somerville08}) because they are not well constrained by observations. This is due to the fact that  they occur within the inner few hundred parsecs, which cannot be  spatially resolved at $z\ge2$ where the co-evolution of galaxies and SMBH largely occurs. It is nearby galaxies that offer the only opportunity to test in detail the prescriptions used in models of galaxy and BH co-evolution.


Constraining the processes of feeding and feedback occurring in the inner few hundred parsecs of nearby active galaxies is the goal of my research group -- called  {\it AGNIFS (AGN Integral Field Spectroscopy)}. We have been doing this via the mapping of the gas kinematics around nearby AGN at spatial resolutions of a few to tens of parsecs, characterizing inflows and outflows. In addition, we have been able, in a few cases, to map the stellar population, looking for the contribution of recent episodes of star formation which can trace the growth of the galaxy bulge. 

In the present contribution, I summarize the most recent results of our AGNIFS group, pointing out similarities to what has been observed in the nuclear region of the Milky Way, even though {\it the MW nucleus is not an AGN}, as pointed out by Deokkeun An, in his contribution to this volume.

\section{Observations}

We have based our studies on integral field spectroscopy (IFS) at the Gemini telescopes. In the optical, we have used the Integral Field Unit of the Gemini Multi-Object Spectrograph (GMOS-IFU), which has a field-of-view of 3.5\arcsec\,$\times$\,5\arcsec\ in one-slit mode or 5\arcsec\,$\times$\,7\arcsec\ in two-slit mode at a sampling of $0.\!''2$ and angular resolution (dictated by the seeing) of $0\!''6$, on average. The resolving power is $R \approx 2500$.

In the near-infrared, we have used the Near-Infrared Integral Field Spectrograph (NIFS) together with the adaptative optics module ALTAIR (ALTtitude conjugate Adaptive optics for the InfraRed), which delivers an angular resolution of $\sim\!0.\!''1$. The field-of-view is 3\arcsec\,$\times$\,3\arcsec\ at a sampling of $0.\!''04 \times 0.\!''1$ and the spectral resolution is $R\approx 5300$ at the Z, J, H and K bands.

\section{Feeding}

There is still no consensus on the mechanisms responsible for transferring mass from galactic scales down to nuclear scales to feed  the SMBH. Theoretical studies and simulations (e.g. \cite[Emsellem et al. 2006]{emsellem06}) have shown how galactic bars efficiently promote gas inflow, but which seems to stall at the Inner Lindblad Ressonance (ILR), located within $\approx$\,1\,kpc from the nucleus forming a circumnuclear ring where star formation is triggered. Inside the inner kiloparsec, \cite{gb05} show that in some galaxies stellar gravity torques can drive gas inwards towards the nucleus. Observations of the inner kiloparsec show also nuclear bars,  which can drive gas to the SMBH. Nevertheless, small-scale disks and nuclear spiral arms are observed more frequently in the inner kiloparsec of active galaxies  than nuclear bars (e.g. \cite[Malkan et  al. 1998]{malkan98}; \cite[Pogge \& Martini 2002]{pogge02}; \cite[Peeples \& Martini 2006]{peeples06}). In particular, we, in \cite{sl07} found that the presence of nuclear dust structures is strongly correlated with activity in early type galaxies. The dust often exhibits a disk and/or spiral morphology, and as dust is an effective tracer of cold, molecular gas, this correlation suggests that these structures map feeding channels to the active nucleus. Recent models indeed support inflows along nuclear spiral arms in active galaxies, such as those of \cite{maciejewski04,hq10,pf12}.

In order to test the hypothesis that nuclear disks and spirals are indeed feeding channels to the AGN, we have been measuring the gas kinematics along these structures. I summarize below the relevant results we have observed so far.

Our first observations of non-circular motions along dusty nuclear spirals was obtained via GMOS-IFU observations of  the inner kiloparsec of NGC\,1097 in \cite{fathi06}, which has a LINER/Seyfert 1 nucleus, by measuring the H$\alpha$+[NII]$\lambda$6584A gas kinematics in the optical. The presence of inflows along the nuclear spirals also in molecular gas  was later reported by \cite{davies09} and more recently via ALMA (Atacama Large Milllimetric Array) observations by \cite{fathi13}. We have next found similar non-circular motions along nuclear spirals -- also from H$\alpha$+[NII] kinematics obtained using GMOS-IFU, around another LINER nucleus, that of NGC\,6951 in \cite{sb07}.  

In M\,81 (LINER/Seyfert 1), using again GMOS-IFU observations, we  (\cite[Schnorr M\"uller et al. 2011]{sm11}) have observed rotation in the stellar velocity field within the inner 100\,pc radius, but a totally distinct kinematics for the gas, which shows inflows along the galaxy minor axis that seem to correlate again with a nuclear spiral. More recently, observations of the inner kiloparsec of  another LINER/Sefyert 1 nucleus, NGC\,7213 (\cite[Schnorr M\"uller et al. 2013b]{sm13b}), have shown a similar result: the ionized gas emission shows a `distorted" rotation pattern. The ``distortions" in the gas velocity field are clearly correlated with a nuclear spiral seen in the structure map, and the residuals -- showing mostly blueshifts in the far side of the galaxy, and redshifts in the near side, can be interpreted as inflows towards the center.
We have measured the mass inflow rate at a distance of $\approx$\,100\,pc from the nucleus as $\approx$\,0.2\,M$_\odot$\,yr$^{-1}$. One interesting result in this case is that we could measure the mass inflow rate as a function of distance from  the nucleus and the largest value is observed $\approx$\,400\,pc from the nucleus, decreasing inwards. This result suggests that the gas is accumulating between 100\,pc and 400\,pc from the nucleus. An speculation is that this accumulated gas may give rise to the formation of new stars in the near future.

All the above discussed cases have LINER, low-luminosity active nuclei. We found that it is indeed easier to ``see" the inflows in galaxies where there is no powerful outflows, as the two may appear superimposed in the line-of-sight velocity field. More recently, we have tried to look for inflows in more luminous targets, such as NGC\,2110 (\cite[Schnorr M\"uller et al. 2013a]{sm13a}). In this galaxy, we could isolate the inflows because it was possible to separate the kinematics in 4 distinct components. One ``cold" component, which is in rotation in the galaxy plane, shows deviation from simple circular rotation, presenting again residual blueshifts in the far side of the galaxy and redshifts in the near side (when a circular velocity field is subtracted), which can be interpreted as inflow towards the center.

The nuclear spirals we have been discussing are dusty, and  should thus be associated with molecular gas. Using the instrument Gemini NIFS, we began also to study the molecular  gas kinematics via observations in the K band. As the spectral resolution of NIFS is almost three times higher than that of the GMOS IFU,  we were able to obtain the gas kinematics using channel maps along the H$_2\lambda2.122\mu$m  emission line profile. The first galaxy in which we found evidence of inflows, also along nuclear spirals, was NGC\,4051 (\cite[Riffel et al. 2008]{riffel08}). We found mostly blueshifts in a spiral arm in the far side of the galaxy, and mostly redshifts in the near side, that can be interpreted as inflows towards the galaxy center if we assume again that the gas is in the plane of the galaxy. 

The H$_2$ emission in the K band maps the ``warm" molecular gas (temperature T$\sim$2000K) which is probably only the ``skin" of  a much larger cold molecular gas reservoir which emits in the milimetric wavelength range (and should thus be observable with the {\it Atacama Large Milimetric Array -- ALMA}, for example). In fact, the warm H$_2$ mass inflow rate within the inner 100\,pc of NGC\,4051 is of the order of only 10$^{-5}$\,M$_\odot$\,yr$^{-1}$. But previous observations of both the warm and cold  molecular gas in a sample of AGN host galaxies show  typical ratios cold/warm H$_2$ masses ranging between 10$^5$ and 10$^7$ as obtained by \cite{dale05}. Applying this ratio to NGC\,4051 leads to a total gas mass inflow rate of $\ge$\,1M$_\odot$\,yr$^{-1}$. 

\begin{figure}
\includegraphics[scale=0.6]{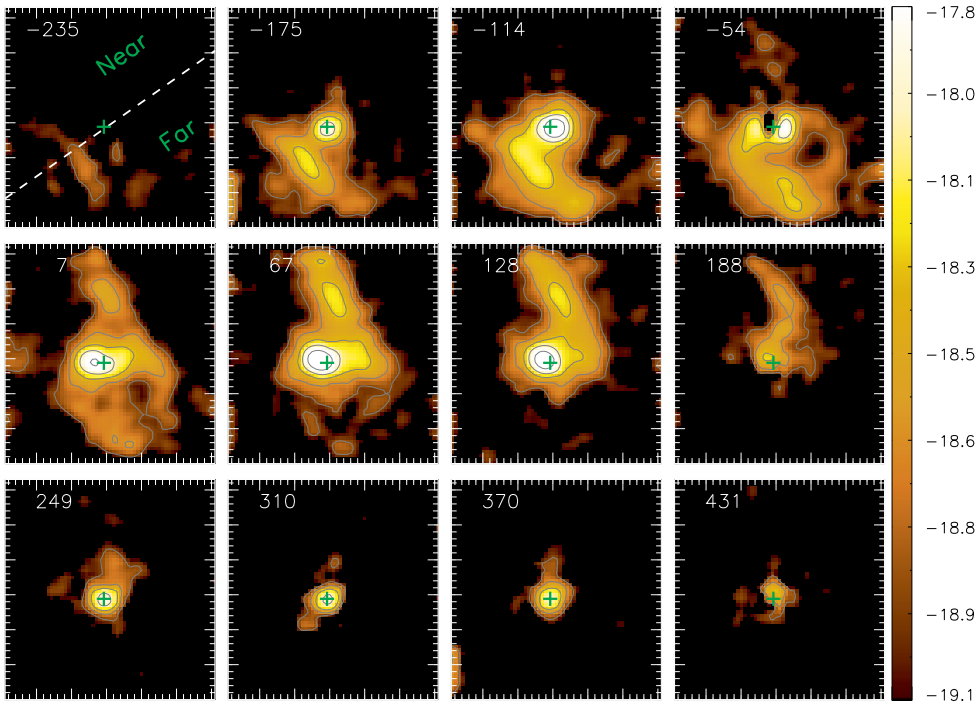}
\caption{Channel maps of the nuclear region of Mrk\,79 along the H$_2\lambda2.122\mu$m emission line profile and centered at the velocity shown in the top-left corner of each panel, for a velocity bin of 60\,\kms. The dashed line shows the major axis of the galaxy, and the near and far sides of the galaxy plane are identified. The large tick marks are separated by 1\arcsec (455\,pc at the galaxy) and the angular resolution of the data is $0.\!''12$ ($\approx$50\,pc).}
\label{mrk79}
\end{figure}

Similar inflows in warm H$_2$ gas were observed in the inner 350\,pc radius of the Seyfert 2 galaxy Mrk\,1066 at 35\,pc spatial resolution (\cite[Riffel et al. 2011a]{riffel11a}) along spiral arms which seem to feed a compact rotating disk with a 70\,pc radius. The mass of warm H$_2$ gas is estimated as 3300\,M$_\odot$, that, corrected to account for the cold component, would imply a reservoir of at least 10$^8$\,M$_\odot$ of cold molecular gas. The corrected mass inflow rate is 0.6\,M$_\odot$\,yr$^{-1}$. More recently, in \cite{riffel13} we have observed inflows also along nuclear spiral arms in the inner 500\,pc of the Seyfert 1 galaxy Mrk\,79, as illustrated in Fig.\,\ref{mrk79}. These channel maps show the flux distributions at the velocities indicated in each panel. The dashed line in the figure shows the major axis of the galaxy, where the near and far side of the galaxy are also identified. We found blueshifts along the spiral arm in the far side and similar redshifts in the near side supporting inflow towards the center.

\subsection{Comparison with the Milky Way (MW)}

In \cite{sl07}  we found a strong correlation between the presence of nuclear dusty spirals and nuclear activity. The MW nucleus cannot be considered active presently; thus we would not expect to see a nuclear dusty spiral. Yet, within the inner 3\,pc of the MW, a small three-armed spiral is seen, as discussed by Lacy in this volume. I  speculate that -- in analogy to what we have found for nearby active galaxies -- that these nuclear spirals in the MW  indicate that very mild  inflows along these spirals could be occurring, leading to a VERY weak LINER, whose mild activity we see only because we are so close. Lacy estimates an inflow rate lower than 10$^{-3}$\,M$_\odot$\,yr$^{-1}$ and point out that the spiral  is co-planar to the so-called molecular circumnuclear disk (CND), argued by \cite{liu12} to be a possible source of material to feed gas towards the center.

\section{Feedback}

The clearest signatures of feedback from AGN are the usually collimated outflows observed in radio (the radio jets) or the broader outflows observed in emission-line kinematics of the the narrow-line region (NLR) as in  \cite{sb92, schmitt94,das06, barbosa09}. 

In the near-IR, while the H$_2$ gas kinematics is dominated by circular rotation and/or inflows in the plane of the galaxy (at  velocities of 50--100\,\kms), the ionized gas emission usually shows outflows (velocities of up to 1000\,\kms) combined with circular rotation (with typical velocities of 200\,\kms). The H$^+$ line-emitting gas (such as in Pa$\beta$) is frequently dominated by rotation, while the  [Fe\,II]$\lambda1.644\mu$m and [Fe\,II]$\lambda1.257\mu$m  emitting gas is frequently dominated by the outflows.  We have thus concluded that the best emission lines to trace outflows are those of [Fe\,II].

NIFS observations of the inner 560\,pc$\times$200\,pc of the Seyfert 1.5 galaxy NGC\,4151 at $\approx$\,7\,pc spatial resolution in \cite{sb10} show that, while the H$_2$ gas is in rotation with velocities lower than 100\,\kms within $\approx$\,50\,pc from the nucleus,  the [Fe\,II]$\lambda1.644\mu$m emitting gas shows a hollow ``conical" outflow with velocities of up to 800\,\kms. From the observed velocities and inferred geometry, we were able to estimate the mass outflow rate: $\approx$\,2\,M$_\odot$\,yr$ ^{-1}$. We were also able to obtain the kinetic power of the outflow which is only $\approx$0.3\% of the bolometric luminosity of the AGN.

In the Seyfert 2 galaxy Mrk\,1066, we (\cite[Riffel et al. 2001a]{riffel11a}) found that the [Fe\,II]$\lambda$1.644$\mu$m emitting gas is collimated along a nuclear radio jet  and reaches outflow velocities of up to 500 \,km\,s$^{-1}$. From the velocity field and geometry of the outflow we estimate a mass outflow rate in ionized gas of  $\approx$\,0.5\,M$_\odot$\,yr$^{-1}$, a value which is curiously of the same order as that of the mass inflow rate in H$_2$ in this galaxy. In a similar study of the Seyfert 2 galaxy Mrk\,1157, we (\cite[Riffel et al. 2011b]{riffel11b}) have obtained a mass outflow rate of $\approx$\,8\,M$_\odot$\,yr$^{-1}$, while in Mrk\,79, we (\cite[Riffel et al. 2013]{riffel13})  have obtained a mass outflow rate of $\approx$\,4\,M$_\odot$\,yr$^{-1}$.

\begin{figure}
\includegraphics[scale=0.9]{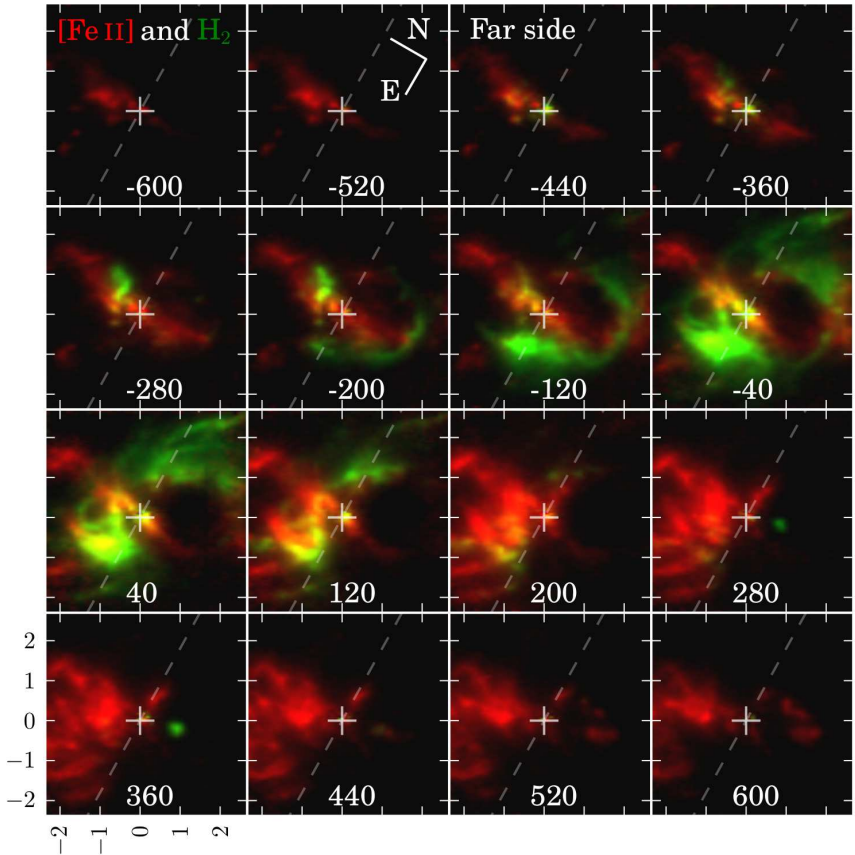}
\caption{Channel maps of the inner 200\,pc (radius) of NGC\,1068 in the [Fe\,II]$\lambda1.644\mu$m (red) and in the H$_2\lambda2.122\mu$m (green) emission lines. The numbers correspond to the central channel velocities in \kms. The nucleus is identified by a cross while the galaxy major axis is identified by a dashed line.}
\label{n1068}
\end{figure}

Recently, we have used NIFS observations of the ``prototypical" Seyfert 2 galaxy NGC\,1068 to map the kinematics of the different gas phases, in \cite{barbosa14}. In Fig.\ref{n1068} we show channel maps of the inner 200\,pc radius  in the [Fe\,II]$\lambda1.64\mu$m (in red) and H$_2\lambda2.122\mu$m (in green) emission lines, at a spatial resolution of 7\,pc. The total flux distribution in [Fe\,II] emission shows an  hourglass structure, while in the channel maps of Fig.\ref{n1068} it shows an ``$\alpha$-shaped" structure in the blueshifted channels and a ``fan-shaped" structure in the redshifted channels. We attribute the blueshifted emission to the front part of the gas outflow modeled by \cite{das06}, while the redshifted emission is attributed to the back part of the outflow. We note that the fan-shaped structure is very similar to that observed in Planetary Nebulae (e.g. NGC\,6302), suggesting a similar mechanism for the origin of the outflow.  Using the inferred geometry and velocity field, we have calculate a mass outflow rate of 6\,M$_\odot$\,yr$ ^{-1}$.

Fig.\,\ref{n1068} also shows that the H$_2$ flux distribution is completely distinct from that of [Fe\,II], as observed in previous cases, presenting a ring-like (radius\,$\approx$\,100\,pc) morphology. The H$_2$ kinematics shows again much smaller velocities than those observed in [Fe\,II] and in common with other active galaxies shows also rotation. Detailed analysis of the H$_2$ kinematics show in addition expansion in the ring in the plane of the galaxy: Fig.\ref{n1068} shows that the H$_2$ emission is blueshifted in the near side of the galaxy and redshifted in the far side. If the gas is in the plane of the galaxy, this implies expansion of the ring.  A detailed analysis of this kinematics has revealed in addition, that the expansion is decelerated from the inner towards the outer border of the ring.

\subsection{Comparison with the MW}

The contribution by McClure-Griffiths to these proceedings reports outflows from the MW center observed in atomic hydrogen (HI) at typical velocities of $\sim$\,200\,\kms. These outflows are not attributed to nuclear activity in the MW center but to winds from young stars formed in the 100\,pc ring (see discussion below), which has an estimated star-formation rate of $\approx$\,0.1\,M$_\odot$\,yr$^{-1}$.

Another signature of feedback from the MW nucleus are the so-called Fermi bubbles that extend to $\approx$\,20\,kpc from the nucleus. Their origin is attributed to an accretion event onto SgrA* or a starburst that occurred there a few 10$^6$\,yr ago, as discussed by Finkbeiner, Conti, Ruszkowski, Lacki, in this volume. I note that the shape of the bubbles at their ``base" (close to the nucleus) are similar to that of the outflow seen in [Fe\,II] emission in NGC\,1068 (Fig.\,\ref{n1068}).

\section{Stellar Population}

\begin{figure}
\includegraphics[scale=0.44]{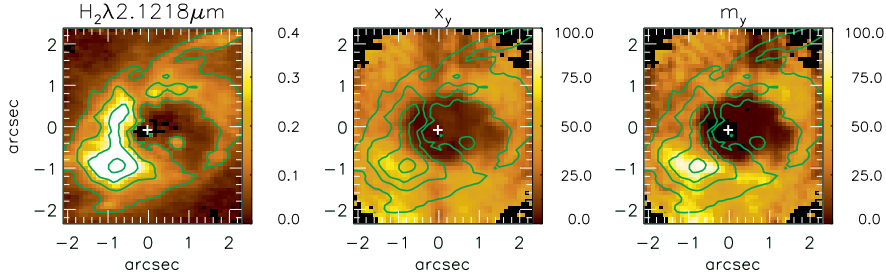}
\caption{Left panel: H$_2$ flux map of the inner 200\,pc radius of the galaxy NGC\,1068. Central panel: contours of the H$_2$ flux map overploted on the map of the percent contribution of the 30\,Myr age stellar population to the total light at 2.1$\mu$m. Right panel: percent contribution of the 30\,Myr age component to the total stellar mass. From \cite{sb12}.}
\label{pop_1068}
\end{figure}

Using the NIFS observations of Mrk\,1066, we \cite{riffel10} published the first resolved two-dimensional  stellar population study of an AGN host  in the near-IR.  Using spectral synthesis, we mapped a 300\,pc circumnuclear ring of intermediated age  (500\,Myr) stellar population which is correlated with a ring of low stellar velocity dispersions. This result has been interpreted as due to an event which led to the capture of gas to the nuclear region with enough gas mass to trigger the formation of new stars 500\,Myr ago. The low velocity dispersion indicates that these stars still keep the ``cold" kinematics of the gas from which they were formed. In \cite{rogerio11}, a similar study revealed almost the same result for the Seyfert\,2 galaxy, Mrk\,1157: another nuclear ring with low stellar velocity dispersion associated to intermediate age stars.  Low stellar velocity dispersion rings were also found in a study of the gas and stellar kinematics of the inner kiloparsec of six nearby Seyfert galaxies (\cite[Barbosa et al. 2006]{barbosa06} using the GMOS IFU, indicating also the presence of stars with ``cold" kinematics, probably also associated with intermediate age stars. 

 More recently, in \cite{sb12}, we have found a smaller (100\,pc radius) ring of young stars in NGC\,1068, dominated by ages of $\approx$\,30\,Myr, as illustrated in Fig.\,\ref{pop_1068}. This ring seems to be correlated with the molecular (H$_2$) ring described above, as shown  by the green contours in the figure, suggesting that the young stars have formed from the gas in the ring.

\subsection{Comparison with the MW}

\begin{figure}
\includegraphics[scale=0.5]{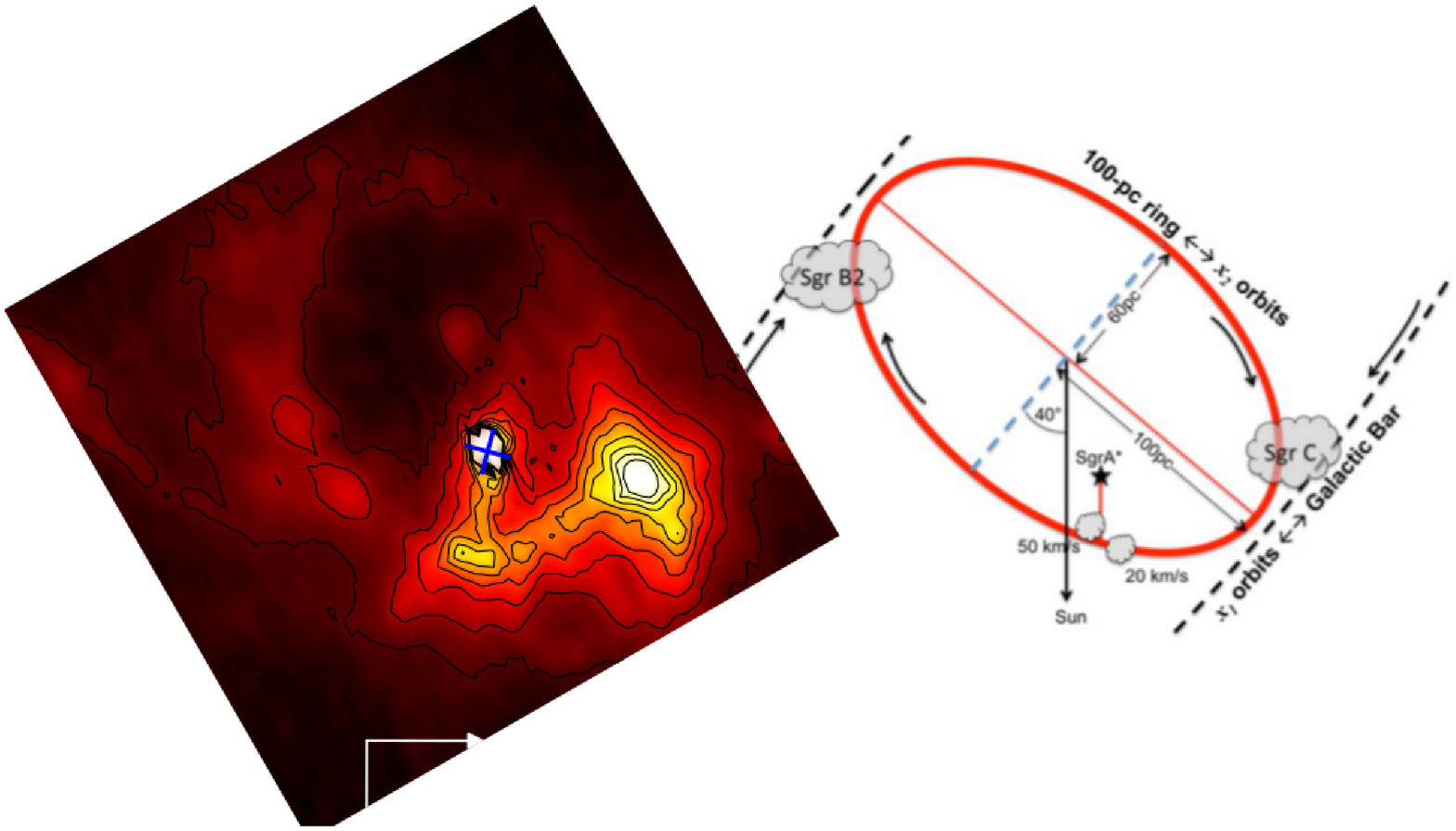}
\caption{Two off-centered 100\,pc molecular rings. Left panel: the H$_2$ ring in NGC\,1068, observed in the near-IR, from \cite{sb12}. Right panel: an sketch of the molecular ring discovered by \cite{molinari11} in Herschell observations of the MW central region.}
\label{rings}
\end{figure}

Using the Herschell satellite, \cite{molinari11} discovered an off-centered dense elliptical ring around the MW center, with a semi-major axis of $\approx$\,100\,pc (see also the contribution by John Bally to this volume). We show in the right panel of Fig.\,\ref{rings} a sketch of this ring -- from the work of \cite{molinari11}, together with an image of the NGC\,1068 molecular ring observed with NIFS in the H$_2\lambda2.122\mu$m emission line. Although the molecular ring in NGC\,1068 seems somewhat rounder than that in the MW, it is also off-centered relative to the nucleus and has a 100\,pc radius, as in the MW. 

It has been pointed out  in this Conference (e.g. in the contributions by Galagher, Longmore, Sakamoto), that the observed star formation rate in the central region of the galaxy is too small for the amount of gas available. We speculate that it may be that the gas in the ring has been accumulating in the ring and has just begun to form stars, and that the star formation rate will increase in the near future in this ring.


\section{Summary and Conclusions}

\noindent{\underline{\it Feeding:} Our observations of the kinematics of both ionized and warm molecular gas in the inner few 100\,pc of nearby AGNs reveal rotation in compact disks and inflows along nuclear spirals, with velocities in the range  50--100\,\kms and mass flow rates  from 0.01 to $\sim$\,1\,M$_\odot$\,yr$ ^{-1}$, observed mainly in LINERs. In \cite{sl07} we showed a clear dichotomy between the nuclear region of early-type AGN hosts -- which always show excess of dust -- and that of non-AGN, supporting the hypothesis that the nuclear spirals and filaments are a necessary condition for the presence of nuclear activity. Nevertheless, the mass accretion rate necessary to feed the AGN -- of $\approx$\,10$^{-3}\,$\,M$_\odot$\,yr$ ^{-1}$, is much smaller than the above mass inflow rate.  At 1\,M$_\odot$\,yr$ ^{-1}$, in 10$^8$\,yr (estimated duration of an activity cycle), $\approx$ 10$^8$\,M$_\odot$ of gas will be accumulated in the inner few hundred parsecs. This  mass  is in agreement with that derived by \cite{martini13}, in a recent study in which we have used Spitzer photometry of  the \cite{sl07} sample to obtain dust masses of the nuclear spirals. The  accumulation of this mass in the nuclear region will probably lead in the near future to the formation of new stars in the galaxy bulge. 
\medskip

\noindent{\underline{\it Stellar population:} Signatures of recent star formation in the bulge of AGN hosts have indeed been  seen, in the form of rings at 100\,pc scales with significant contribution from stars of ages in the range 30\,Myr$\leq$age$\leq$\,700\,Myr. These results suggest that we are witnessing the co-evolution of the SMBH and their host galaxies in the near Universe: while the SMBH at the center grows at typical rates of 10$^{-3}$\,M$_\odot$\,yr$ ^{-1}$, the bulge growths at typical rates of 0.1--1\,M$_\odot$\,yr$ ^{-1}$. A similar evolution scenario has been previously proposed by \cite{sb01}.

\medskip

\noindent{\underline{\it  Feedback:}  Our IFS observations usually reveal ionized gas outflows in nearby AGN, not so frequently in LINERS, but always observed in Seyfert galaxies. They extend to a few hundred of parsecs from the nucleus,  are oriented at random angles to the galaxy plane, and reach velocities in the range 200 --1000\,\kms. The total ionized gas mass in these outflows are $\sim$\,10$^{6-7}$\,M$_\odot$, and the mass outflow rates are in the range 0.5--10\,M$_\odot$\,yr$ ^{-1}$, which is similar to the range of the mass inflow rates, although only in a couple of galaxies we observe both inflows and outflows (e.g. in Mrk\,1066). The fact that the mass outflow rates are 100 -- 1000 times the AGN accretion rate supports the idea that the observed outflows are due to mass loading of an AGN outflow (which should be at most equal to the AGN accretion rate) as it moves through the circumnuclear interstellar medium of the host galaxy.

\medskip

\noindent{\underline{\it Comparison with the MW:} Our galaxy does not have an active nucleus. Its very mild nuclear activity can only be observed because the MW nucleus is about 1000 times closer than the closest AGNs. Yet, there are some similarities with what we see around nearby AGNs: the presence of a compact molecular disk and nuclear spiral; the presence of a molecular ring at 100\,pc from the nucleus, and the presence of past energetic outflows that can be attributed to a past ejection event associated to SgrA*.

In order to connect episodes of nuclear activity in nearby galaxies and the non-active phase of the MW, I shall use a speculative but educated evolutionary scenario that I proposed some time ago in \cite{sb01}, as follows. The nuclear activity is triggered by mass inflows towards the central region of the galaxies. The observed  mass inflow rate we have measured in LINERs (of the order of one to a few M$_\odot$\,yr$^{-1}$) are much larger than the accretion rate necessary to power the AGN. This gas is thus probably accumulated in the inner few hundred pcs giving origin to the observed nuclear disks and spirals in the hosts, and are probably later consumed by episodes of star formation. These episodes are followed by the increase of the nuclear activity. This evolution is supported by the fact that, in the majority of active galaxies, the circumuclear rings are dominated by stars of intermediate age ($\sim$10$^{7-8}$ yr) rather than young ($\sim$10$^6$\,yr), consistent with the idea that, in most AGN the nuclear activity is observed after the episodes of star formation. In this scenario, the MW, which has a molecular ring at 100\,pc and the CND within the inner 3\,pc, as well as a population of young stars in the nuclear region, may have just begun the phase of starburst activity in the nuclear region, which will be followed, probably tens of millions of years into the future, by the triggering of nuclear activity in SgrA*.




\end{document}